# Crystal Field Effects in CeIrIn$_5$

A.D. Christianson[1], J.M. Lawrence[1], P.S. Riseborough[2], N.O. Moreno[3], P.G. Pagliuso[3], E.D. Bauer[3], J.L. Sarrao[3], W. Bao[3], E.A. Goremychkin[4], S. Kern[5], F.R. Trouw[3], and M.P. Hehlen[3]

[1] University of California, Irvine, California 92697
[2] Temple University, Philadelphia, Pennsylvania 19122
[3] Los Alamos National Laboratory, Los Alamos, New Mexico 87545
[4] Argonne National Laboratory, Argonne, Illinois 60439
[5] Colorado State University, Fort Collins, Colorado 80523

Mailing Address: Los Alamos National Laboratory, MS K764, Los Alamos, NM 87545
USA
Fax: 505-665-7652
Email: achristianson@lanl.gov

In this work, we study crystalline electric field effects in the heavy fermion superconductor CeIrIn$_5$. We observe two regions of broad magnetic response in the inelastic neutron scattering spectra at 10 K. The first corresponds to the transition between the $\Gamma_7$ groundstate doublet and the first excited state doublet at 4 meV interwoven with a broad quasielastic contribution. The second region corresponds to the transition between the ground state and the second excited state doublet at 28 meV. The large Lorentzian half-widths of the peaks (~10 meV) calls into question calculations for the specific heat and magnetic susceptibility that assume sharp crystal field levels. Consequently, we have calculated the inelastic neutron scattering spectra and magnetic susceptibility using the Anderson impurity model within the non-crossing approximation (NCA) including the effects of crystal field level splitting.

Keywords: inelastic neutron scattering, heavy fermions, Kondo effect, superconductivity, crystal field splitting

The discovery of the CeMIn$_5$ (M = Co, Rh, Ir)[1,2,3] series of heavy fermion superconductors offers the opportunity to systematically explore heavy fermion superconductivity in closely related materials. Despite numerous attempts, the origin of superconductivity in the CeMIn$_5$ series has yet to be established unambiguously. There have been several proposals put forth, based on both experimental and theoretical considerations, suggesting that crystal field (CF) splitting may play a crucial role in the heavy fermion superconductivity observed in the CeMIn$_5$ series[4,5,6]. Additionally, from the point of view of bulk measurements such as specific heat, magnetic susceptibility, and thermal expansion, there is disagreement about the particular CF parameters in these materials. Therefore a direct measurement of the CF level splitting is desirable. In this work we focus on CF effects in the heavy fermion superconductor CeIrIn$_5$.

Crystal Field level splittings in heavy fermion systems are typically observed as broadened peaks in an inelastic neutron scattering (INS) spectrum. The broadening of the peaks in the INS is due to hybridization between the localized f-electrons and the conduction electrons. The hybridization has the added effect of renormalizing the energy splitting of the CF levels. In the case of CeIrIn$_5$, the CF Hamiltonian is that of Ce$^{3+}$ in a tetragonal environment and may be written as

$$H_{CF} = B_2^0 O_2^0 + B_4^0 O_4^0 + B_4^4 O_4^4 \tag{1}$$

where the $B_l^m$ are the CF parameters and the $O_l^m$ are the Stevens operator equivalents. Diagonalization of this Hamiltonian yields the following wave functions.

$$\Gamma_7^1 = \alpha\left|\pm\frac{5}{2}\right\rangle - \beta\left|\mp\frac{3}{2}\right\rangle \quad \Gamma_7^2 = \beta\left|\pm\frac{5}{2}\right\rangle + \alpha\left|\mp\frac{3}{2}\right\rangle \quad \Gamma_6 = \left|\pm\frac{1}{2}\right\rangle \tag{2}$$

The form of the wave functions are determined by symmetry considerations alone, while the mixing parameters (α and β) and the energy splitting can be determined by fitting to experiment.

The experiments described here were performed on two inelastic chopper spectrometers: PHAROS at the Manuel Lujan Neutron Science Center (Los Alamos National Laboratory) and LRMECS at the Intense Pulsed Neutron Source (Argonne National Laboratory). Approximately 50 g of CeIrIn$_5$ and 40 g of LaIrIn$_5$ single crystals were produced by growth in indium flux and powdered for these investigations. Data were collected at incident energies (E$_i$) of 30 and 60 meV at 10 K and at elevated temperatures (not shown). The counting time for each incident energy at each temperature was approximately two days. The effect of the strong neutron absorption of In and Ir was minimized by performing the experiments in a uniform flat plate sample geometry, which not only maximized the amount of sample in the beam while keeping the neutron path length as short as possible, but also allowed for an accurate absorption correction. The magnetic portion of the INS spectrum has been determined by S$_{mag}$ = S(CeIrIn$_5$) – f S(LaIrIn$_5$), where f = 0.83 is the ratio of the total scattering cross sections of CeIrIn$_5$ and LaIrIn$_5$. Other methods of subtraction of the nonmagnetic scattering give similar results. The dependence of S$_{mag}$ on the magnetic form factor has been removed in the INS spectra shown below.

Figure 1(a) and (b) displays S$_{mag}$ determined from data recorded on LRMECS at 10 K with an E$_i$ = 30 meV and 60 meV respectively. Figure 1(c) displays similar data from PHAROS at 18 K with E$_i$ = 30.2 meV. The most prominent feature of these data is that the CF levels are extremely broad. The CF peaks are broadened to such a degree that we are unable to resolve the quasielastic scattering, despite performing higher resolution experiments on PHAROS with a FWHM at the elastic line of 1.4 meV (Figure 1(c)). Figure 1(a-c) also display fits to the INS spectra. The fits contain an overall scale factor for each dataset, the CF parameters and a single Lorentzian half-width $\Gamma_{IE}$ of the CF excitations at each temperature. To prevent proliferation of fitting parameters the quasielastic width $\Gamma_{QE}$ has been constrained to be 1/4 $\Gamma_{IE}$. For CeRhIn$_5$[7] and CeCoIn$_5$[8] this gives values of $\Gamma_{QE}$ that are in good agreement with estimates from NMR experiments[9]. The resulting crystal field parameters for CeIrIn$_5$ are $B_2^0 = -1.21$ meV, $B_4^0 = 0.063$ meV, and $B_4^4 = 0.075$ meV. This implies that the groundstate ($\Gamma_7^1$) - first excited state ($\Gamma_7^2$) splitting is $E_1 = 4$ meV[10], while the ground state - second excited state ($\Gamma_6$) splitting is $E_2 = 28$ meV. Furthermore, the admixture of the $J_z = 5/2$ and 3/2 states is nearly equal with $\beta = 0.76$. Finally, the half-width of the inelastic excitations is indeed large and is found to be $\Gamma_{IE} = 10$ meV, indicating large f-conduction electron hybridization.

Such large hybridization suggests that Kondo physics is important in CeIrIn$_5$. Figure 2(a) shows calculations for the INS spectra of an Anderson impurity subject to crystal field splitting within the non-crossing approximation (NCA) assuming the hybridization (V) = 470 meV and figure 2(b) shows the corresponding calculation of the magnetic susceptibility. Since $\Gamma_{IE} > E_1$, the theoretical result is insensitive to $E_1$ and we obtain good fits by assuming a ground state quartet; the degeneracy of the ground state is, in effect, larger than 2 in CeIrIn$_5$ due to the f-conduction electron hybridization. We assume a bare crystal field level at 22.6 meV and an equal admixture of the $J_z = 5/2$ and 3/2 states ($\alpha^2=\beta^2=0.5$). The NCA calculations reproduce the magnetic susceptibility reasonably well, with some deficiencies at lower temperatures, which may be in part due to the onset of magnetic correlations.

It is instructive to compare these results to those for two other ambient pressure heavy fermion superconductors. The separation $E_1$ between the ground state and the excited states is 30 meV in $CeCu_2Si_2$[11] and 8.6 meV in $CeCoIn_5$[8] but is 6.9 meV in antiferromagnetic $CeRhIn_5$[7]. The inelastic width $\Gamma_{IE}$ is 6.5 meV in $CeCoIn_5$[8], 10 meV in $CeCu_2Si_2$[11] but is only 2.3 meV in $CeRhIn_5$[7]. Hence while there is no correlation between superconductivity and $E_1$, it does appear that moderately large hybridization widths $\Gamma_{IE}$ are necessary for superconductivity. More detailed comparisons among the members of $CeMIn_5$ will be given in a forthcoming publication[12].

In summary, we have determined the CF level scheme of $CeIrIn_5$ by least squares fitting to a CF model. The CF excitations are quite broad. The resulting crystal field parameters were then used as input to Anderson impurity calculations which reproduce the INS spectra and the magnetic susceptibility. The moderately large width of the CF excitations is related to the f-conduction electron hybridization, which appears to be a condition for superconductivity in this heavy Fermion compound.

Figure Captions:

Figure 1. (a) and (b) symbols are $S_{mag}$ determined from data collected on LRMECS as described in the text for $E_i$ = 30 and 60 meV respectively at T = 10 K. In (c) the symbols are $S_{mag}$ determined from data collected on PHAROS with $E_i$ = 30.2 meV at 18 K. The solid line in (a-c) is a least squares fit as described in the text.

Figure 2. (a) The symbols are the data from figure 1(a) and (b). The solid line is a NCA calculation with the hybridization V = 470 meV. (b) Shows the corresponding NCA calculation (solid line) for the magnetic susceptibility compared to the experimental results (symbols).

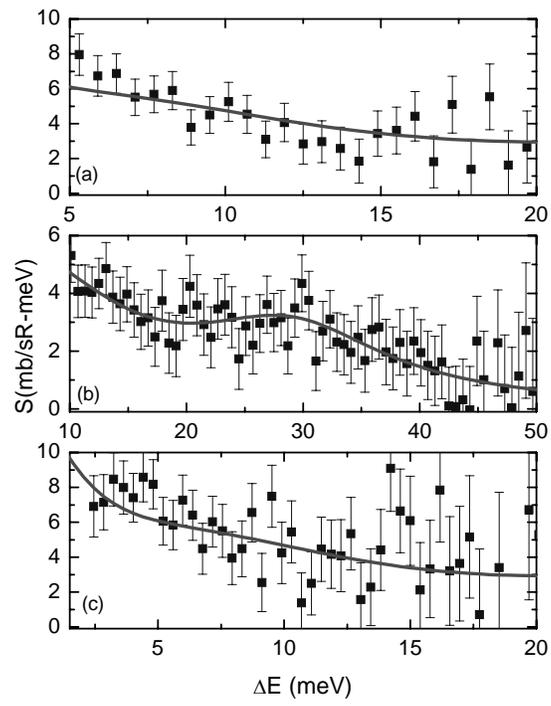

Figure 1: A.D. Christianson *et al.*

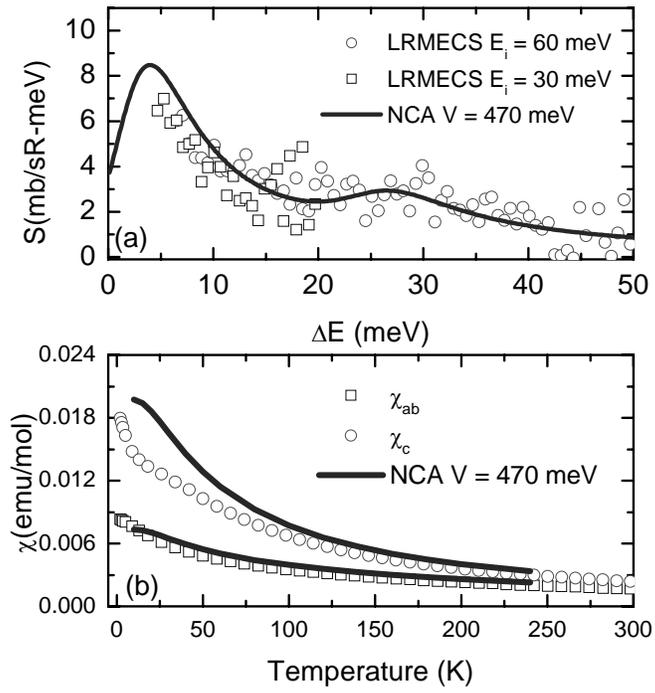

Figure 2 : A.D. Christianson *et al.*